\documentclass[final]{svjour3}
\usepackage{graphicx}
\usepackage{rotating}
\usepackage{amssymb}
\usepackage{mathptmx}
\usepackage{cite}
\usepackage{subfig}
\usepackage{diagbox}
\makeatletter
\journalname{Journal of Low Temperature Physics}

\begin{document}

\newcommand{\hdblarrow}{H\makebox[0.9ex][l]{$\downdownarrows$}-}
\title{Design of a testbed for the study of system interference in space CMB polarimetry}

\author{T. Ghigna\textsuperscript{{\normalfont \textit{a,b}}} \and T. Matsumura\textsuperscript{\normalfont \textit{b}} \and M. Hazumi\textsuperscript{\normalfont \textit{c,b,d,e}} \and S. L. Stever\textsuperscript{\normalfont \textit{b}} \and Y. Sakurai\textsuperscript{\normalfont \textit{b}} \and N. Katayama\textsuperscript{\normalfont \textit{b}} \and \\ A. Suzuki\textsuperscript{\normalfont \textit{f}} \and B. Westbrook\textsuperscript{\normalfont \textit{g}} \and A. T. Lee\textsuperscript{\normalfont \textit{g,f}}}

\institute{\textsuperscript{\textit a}Sub-department of Astrophysics, University of Oxford, Oxford, OX1 3RH, UK, 
\textsuperscript{\textit b}Kavli IPMU (WPI), UTIAS, The University of Tokyo, Kashiwa, Chiba 277-8583, Japan, 
\textsuperscript{\textit c}KEK, Tsukuba, 300-3256, Ibaraki, Japan, 
\textsuperscript{\textit d}SOKENDAI, Hayama, 240-0193, Kanagawa, Japan,
\textsuperscript{\textit e}ISAS-JAXA, Sagamihara, 252-5210, Kanagawa, Japan, 
\textsuperscript{\textit f}LBNL, Berkeley, 94720, California, USA, 
\textsuperscript{\textit g}Department of Physics, University of California Berkeley, Berkeley, 94720, California, USA} 

\maketitle

\begin{abstract}

LiteBIRD is a proposed JAXA satellite mission to measure the CMB B-mode polarization with unprecedented sensitivity ($\sigma_r\sim 0.001$). To achieve this goal, $4676$ state-of-the-art TES bolometers will observe the whole sky for 3 years from L2. These detectors, as well as the SQUID readout, are extremely susceptible to EMI and other instrumental disturbances e.g. static magnetic field and vibration. As a result, careful analysis of the interference between the detector system and the rest of the telescope instruments is essential. This study is particularly important during the early phase of the project, in order to address potential problems before the final assembly of the whole instrument. We report our plan for the preparation of a cryogenic testbed to study the interaction between the detectors and other subsystems, especially a polarization modulator unit consisting of a magnetically-rotating half wave plate. We also present the requirements, current status and preliminary results.

\keywords{CMB, Bolometers, Transition Edge Sensors}

\end{abstract}
\vspace{-5mm}
\section{Introduction}
\vspace{-3mm}
The current main goal of the CMB community is a measurement of the primordial B-mode signal. An accurate measurement will allow us to understand more about the evolution of our Universe and especially about the first moments after the Big Bang (\cite{seljak_zaldarriaga99, zaldarriaga_seljak99, seljak_zaldarriaga97, zaldarriaga_seljak97, kamionkowski97}). Our current theories point in the direction of a very rapid inflationary expansion of the Universe to explain its observed evolution. The best way to probe this theory, as of today, is to measure the polarized CMB signal on large angular scales ($\ell \lesssim$ 200). At these scales the B-mode signal should be dominated by its primordial inflationary component for a value of the tensor-to-scalar ratio parameter $r\gtrsim0.01$. At smaller angular scales, the B-mode signal originating from gravitational lensing of the E-mode signal due to cosmological structures is dominant over the primordial signal, whereas at very large scales ($\ell<10$) the primordial signal should be dominant over the lensing signal for $r\gtrsim 0.001$.
\begin{figure}[htbp]
    \begin{center}
    \includegraphics[width=1\linewidth,keepaspectratio]{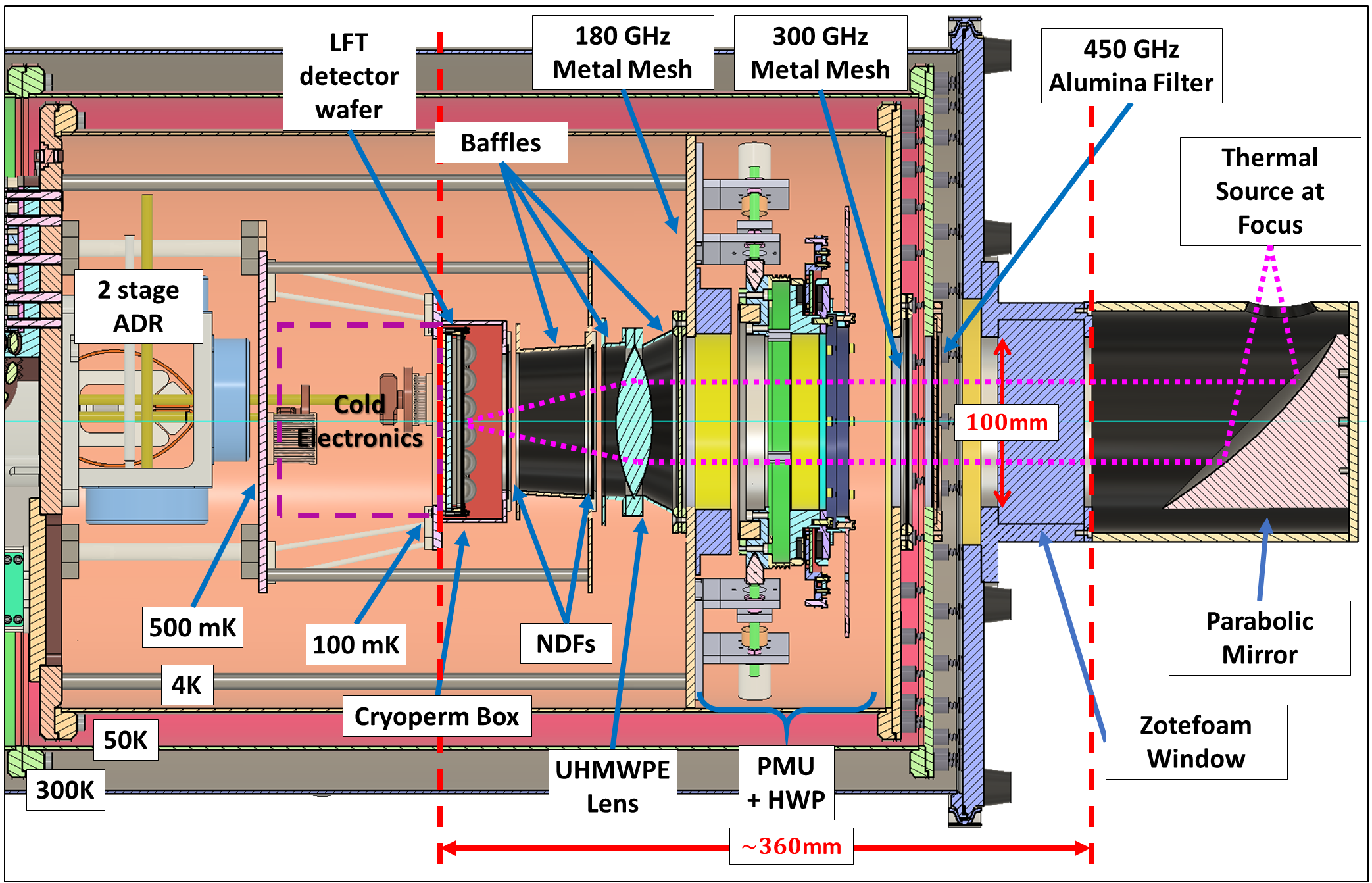}
    \caption{\small{3D model of the cryogenic testbed. (Color figure online.)}}
    \label{fig:testbed}
    \end{center}%
    \vspace{-7mm}
\end{figure}
The only way to access the largest scales and observe both the predicted recombination and reionization bumps of the spectrum is from space, where the absence of atmosphere presents perfect conditions for a large-scale survey of the sky. The LiteBIRD mission (\cite{matsumura14, suzuki18}), consisting of 3 telescopes $-$ Low Frequency Telescope (LFT, $34-161$ $\rm{GHz}$), Mid Frequency Telescope (MFT, $89-224$ $\rm{GHz}$) and High Frequency Telescope (HFT, $166-448$ $\rm{GHz}$) $-$ has been designed with this goal in mind, with 3 years of observation from L2 and $4676$ polarization sensitive bolometric Transition Edge Sensors (TES).

Kavli IPMU is responsible for the development of the LFT Polarization Modulator Unit (PMU) consisting of a continuously-rotating Superconductive Magnetic Bearing (SMB), a cryogenic holder mechanism, a rotation angle encoder, and a multi-layer sapphire Achromatic Half Wave Plate (AHWP) with the required Anti-Reflection Coating (ARC) made by laser-machining sub-wavelength structures (\cite{sakurai18, komatsu18, komatsu19}). Before assembling the full telescope, we plan to adapt an existing Adiabatic Demagnetization Refrigerator (ADR) cryostat to host a cryogenic polarimeter to study the detector performance together with the PMU as part of a full telescope system. The two main challenges for such a system are the extremely low saturation power of space optimized bolometers ($\sim 0.5 $ $\rm{ pW}$) and the limited cooling capacity of an ADR cryostat ($\sim 120 $ $\rm{ mJ}$ at $0.1 $ $\rm{ K}$).

In the following sections, we will describe the testbed design and requirements. We will also show and discuss some of the preliminary results we obtained from single TES detector tests. We are performing these tests in order to build some experience with the TES detectors before using a more complex system, as well as cross-checking detector performances.
\vspace{-5mm}
\section{Testbed cryostat}
\vspace{-3mm}
The main goal of LiteBIRD is measuring the primordial B-mode signal on large angular scales. However, this is impossible without detailed knowledge of the instrument and its potential systematic effects. In Table \ref{tab:tests}, we identify the items to test with the proposed assembly, the potential issues, and the possible sources. Here we list the potential problems
\begin{table}[htbp]
    \centering
    \begin{tabular}{|c||c|c|c|c|c|c|}
    \hline
    \backslashbox{Tests}{Sources} & Pulse tube & ADR & Lens & Filters & PMU \& HWP & Environment \\
    \hline
    \hline
    White noise & V & M & O & O & V, M, O & O, CR, SL \\
    \hline
    1/f noise & T, V & T, M & T & T & T, V & T, CR, SL \\
    \hline
    Correlated noise &  & M &  &  & M, S & CR \\
    \hline
    Spurious signals &  & M & SW & SW & S, SW & SL, CR \\
    \hline
    Time constant(s) &  & M &  &  & S & CR \\
    \hline
    Gain stability & T & T, M & T, O & T, O & T, O & T, CR, O, SL \\
    \hline
    Beam \& Side lobes &  &  & G & G & S, G & SL \\
    \hline
    Bandpass (FTS) &  &  & SW & SW & SW &  \\
    \hline
    I to P leakage &  &  & G & G & S, G &  \\
    \hline
    Polarization Angle &  &  &  &  & S &  \\
    \hline
    \end{tabular}
    \caption{\small{In this table we summarize the test we want to conduct with the proposed set-up, the possible issues and the sources of these possible systematic effects. {\it Legend:} V=vibrations, M=magnetic fields, O=optical loading, S=synchronous signals, SL=stray light, SW=standing waves, CR=cosmic rays, T=thermal instability, G=ghosting.}}
    \label{tab:tests}
    \vspace{-7mm}
\end{table}of the detector system when combined with the other components (optical and mechanical) of the telescope. 

As an example, we have conducted a study to derive the  requirements on the knowledge of the bandpass resolution ($\Delta\nu \sim 0.5 $ $\rm{ GHz}$) and the gain uncertainty at the end of the mission, between $\sim$ few $\%$ and $\sim 0.1 \%$, depending on the frequency band (\cite{ghigna}). Our first goal is to test the possibility of achieving this level of accuracy with a testbed experiment consisting of all major components of the telescope.

The LFT focal plane design consists of a modular assembly of 8 detector wafers (\cite{sekimoto18, sugai19}), each $105\times 105$ $\rm{mm}$, covering a frequency range from $34$ $\rm{GHz}$ to $161$ $\rm{GHz}$ (9 frequency bands) with trichroic pixels. Depending on the frequency range, each wafer has $16$ or $36$ pixels, arranged in a square pattern. With the proposed system we will be able to test at the same time a sub-set of $9$ (for the $16$ pixel wafers) or $16$ pixels (for the $36$ pixel wafers).

The driving requirements for the system in Fig. \ref{fig:testbed}, other than the already mentioned ADR cooling capacity constraints and detector saturation power, are the need for a collimated beam through the HWP to study its optical properties and a static magnetic field (from the HWP magnetic bearing) $<0.2 $ $\rm{ G}$ at the focal plane (see Section \ref{sec:mag} for details).
\vspace{-5mm}
\subsection{Thermal and optical requirements}
\vspace{-3mm}
For our testbed, we plan to modify an existing 2-stage ADR cryostat from HPD (\cite{HPD}) with the following thermal specifications: $\sim 40 $ $\rm{ W}$ of cooling power at the nominal $45 $ $\rm{ K}$ stage, $\sim 1.35 $ $\rm{ W}$ of cooling power at the nominal $4.2 $ $\rm{ K}$ stage (these first 2 stages are cooled by a pulse tube), $1.2 $ $\rm{ J}$ of cooling capacity at the {\it GGG} stage (nominally at $500 $ $\rm{ mK}$) and finally $120 $ $\rm{ mJ}$ of cooling capacity at the {\it FAA} stage (nominally at $100 $ $\rm{ mK}$).

The limited cooling capacity of the ADR requires a careful thermal design to minimize the thermal loading on the $100 $ $\rm{ mK}$ focal plane. 
To do so, we designed a support structure for the $500 $ $\rm{mK}$ stage, consisting of four $100 $ $\rm{ mm}$ Vespel SP1 (\cite{vespel}) shafts with $10 $ $\rm{ mm}$ outer
\begin{table}[htbp]
    \vspace{-5mm}
    \centering
    \begin{tabular}{|c||c|c|c|c|}
    \hline
    Stage & Cooling power/capacity & Radiative loading & Structure loading & Wire loading \\
    \hline
    \hline
    $50 $ $\rm{ K}$ & $40 $ $\rm{ W}$ & $5 $ $\rm{ W}$ & $1 $ $\rm{ W}$ & $5 $ $\rm{ W}$ \\
    \hline
    $4 $ $\rm{ K}$ & $1.35 $ $\rm{ W}$  & $0.6 $ $\rm{ W}$ & $0.1 $ $\rm{ W}$ & $0.2 $ $\rm{ W}$ \\
    \hline
    $500 $ $\rm{ mK}$ & $1.2 $ $\rm{ J}$ & $6 $ $\rm{ \mu W}$ & $10 $ $\rm{ \mu W}$ & $0.1 $ $\rm{ \mu W}$ \\
    \hline
    $100 $ $\rm{ mK}$ & $120 $ $\rm{ mJ}$ & $25 $ $\rm{ nW}$ & $1 $ $\rm{ \mu W}$ & $55 $ $\rm{ nW}$ \\
    \hline
    \end{tabular}
    \caption{\small{Summary of the cooling power (or capacity) and the thermal loading due to radiation, support structures, and wires, for each stage of our dedicated cryostat. This configuration will allow $\sim 10$ hours of continuous operation. The ADR recycling time is $\sim 5$ hours.}}
    \label{tab:thermal}
    \vspace{-5mm}
\end{table}diameter and $9 $ $\rm{ mm}$ inner diameter, while the $100 $ $\rm{ mK}$ stage is supported by a octapod structure made with $100 $ $\rm{ mm}$ Vespel SP22 shafts with a diameter of $3 $ $\rm{ mm}$.
Radiative loading is the other relevant component in our thermal budget. To minimize the IR loading we designed a filter stack consisting of a $450 $ $\rm{ GHz}$ cut-off alumina low-pass filter with a Stycast (\cite{stycast}) anti-reflection coating (\cite{inoue14}) on the $50 $ $\rm{ K}$ stage, and two metal mesh low-pass filters on the $4 $ $\rm{ K}$ stage: one with a $300 $ $\rm{ GHz}$ cut-off before the HWP, and the second after the HWP with a $180 $ $\rm{ GHz}$ cut-off. A complete summary of our thermal calculation and requirements can be seen in Table \ref{tab:thermal}. At the design stage we are not accounting for diffraction effects, we are considering the temperature of all components to be uniform, and the cut-off frequency of the filters to be ideal.

Although this configuration is enough to reduce the radiative loading for thermal purposes, we must also consider the low saturation power of LiteBIRD TES bolometers ($\sim 0.5 $ $\rm{ pW}$). Therefore, we have included in our design two additional filters, one at $500 $ $\rm{ mK}$ and the second at $100 $ $\rm{ mK}$ (in front of the focal plane). These two filters could be NDFs (Neutral Density Filters) or absorptive filters made with a slab of Eccosorb (\cite{eccoCR}), to reject $\sim 99 \%$ of the incoming radiation and reduce the power at the detector to $\sim 0.2 $ $\rm{ pW}$ (a $100 $ $\rm{ GHz}$ band with $30 \%$ bandwidth has been assumed for the calculation). We are still considering which solution is more suitable, but the absorptive filter solution seems more likely due to the suppression of reflections without having to complicate the geometry by tilting one or more of the optical elements.
\vspace{-5mm}
\subsection{Magnetic field requirements}\label{sec:mag}
\begin{figure}[htbp]
    \vspace{-5mm}
    \centering
    \subfloat{{\includegraphics[width=.45\textwidth]{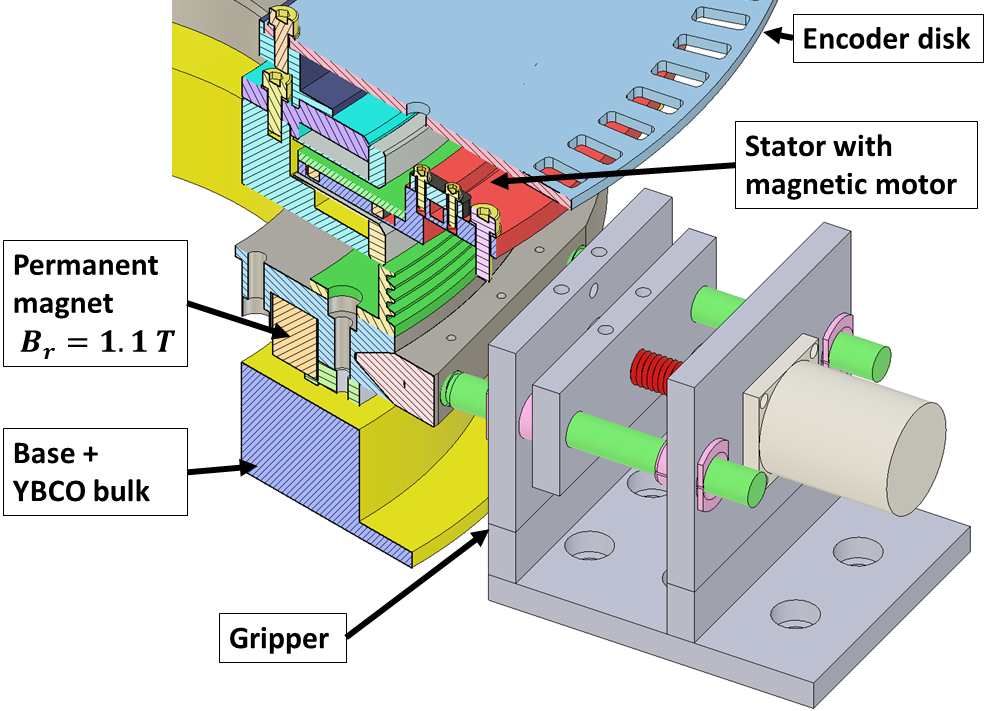} }}%
    \subfloat{{\includegraphics[width=.45\textwidth]{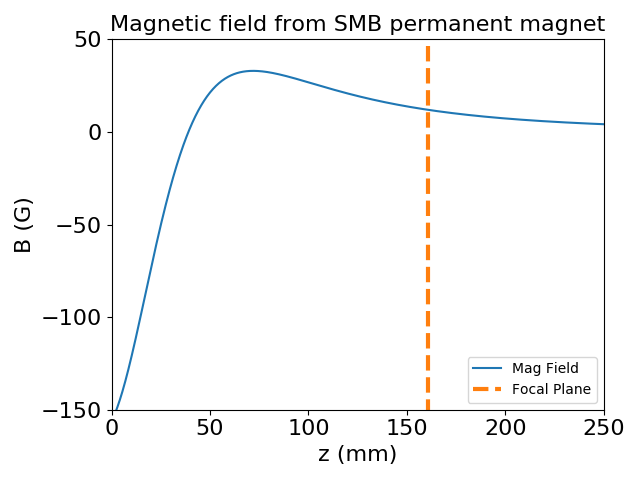} }}%
    \caption{\small{{\it Left:} A cut view of the HWP rotational mechanism, showing the superconductive YBCO bulk and the permanent magnet. {\it Right:} Calculation of the magnetic field at the focal plane level $\sim 160 $ $\rm{ mm}$ away from the permanent magnet. The calculation shows a magnetic field of $\sim 11.8 $ $\rm{ G}$. (Color figure online.)}}%
    \label{fig:HWP}%
    \vspace{-5mm}
\end{figure}

\vspace{-3mm}
A second important item for the definition of the testbed design is the magnetic field due to the SMB of the HWP rotational mechanism (see Fig. \ref{fig:HWP}). We derived a $< 0.2$ $\rm{ G}$ requirement for the magnetic field at the level of the focal plane. External magnetic fields cause the transition temperature of the TES to change (\cite{vavagiakis18}). This change increases the phonon noise (\cite{Mather1982, SuzukiThesis}) of the bolometer.
The requirement is found by imposing a $\sim 10\%$ limit on the degradation of the phonon noise, and assuming the TES transition temperature sensitivity to an external magnetic field of $\sim 4$ mK/G. We find that for LiteBIRD detectors, operating at a bath temperature $T_b\sim$100 mK with a nominal transition temperature $T_c\sim$171 mK, the maximum variation of $T_c$ that we can tolerate is $\sim 7$ mK, which corresponds to an external magnetic filed of $\sim$ 2 G.
However, since the HWP magnetic bearings are not the only magnetic field source, we assign 10\% of the total budget ($\sim 2$ G) to the magnetic bearings: $\sim 0.2$ G.
\begin{figure}[htbp]
    \vspace{-5mm}
    \centering
    \subfloat{{\includegraphics[width=.325\textwidth]{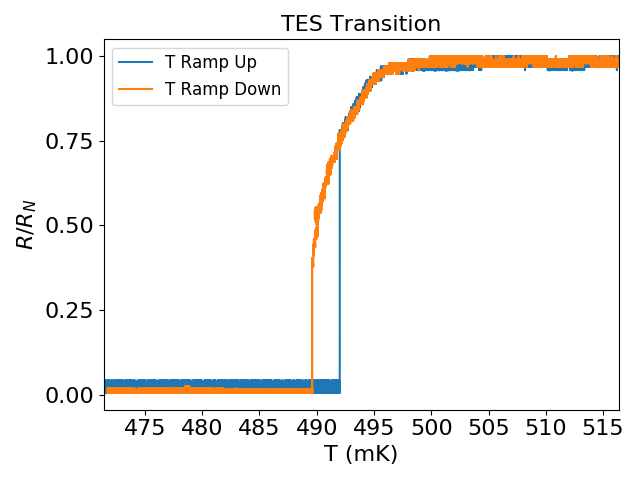} }}%
    \subfloat{{\includegraphics[width=.325\textwidth]{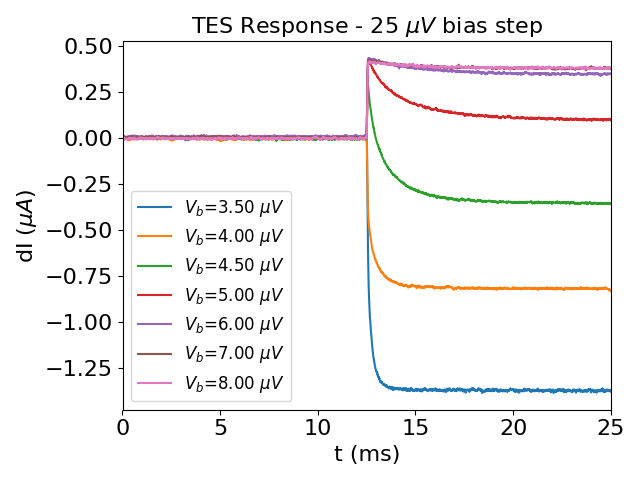} }}%
    \subfloat{{\includegraphics[width=.325\textwidth]{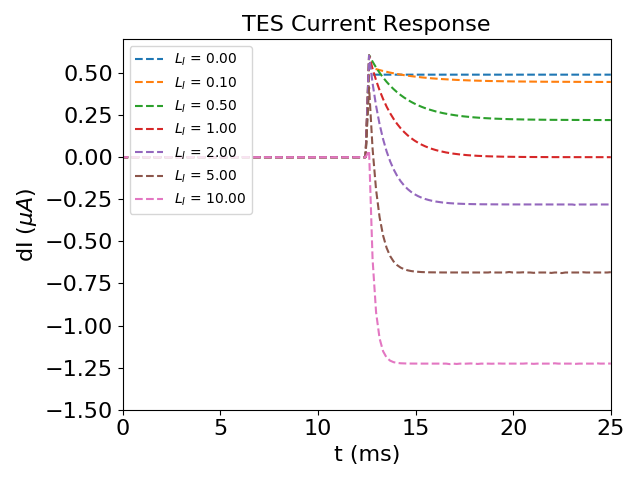} }}%
    \caption{\small{{\it Left:} Transition current ($T_c$) measurement for a TES detector. The measurement has been performed with both the bath temperature ramping up and down. {\it Center:} Detector data from a square wave signal on the bias line for different bias voltages (the bath temperature for these measurement was $\sim 300$ mK). {\it Right:} Simulated response for the same input signal for different loop gain values (bias voltages). The TES steady state output current has been shifted to $0$ $\rm{\mu A}$. (Color figure online.)}}%
    \label{fig:TES1}
    \vspace{-5mm}
\end{figure}

The permanent magnet of the SMB is a ring with outer and inner radii of $68.5 $ $\rm{ mm}$ and $58.5 $ $\rm{ mm}$ (respectively), $12 $ $\rm{ mm}$ thickness, and presents a remanence magnetization $B_r = 1.1\times 10^4 $ $\rm{ G}$. This geometry results in a magnetic field at the center of the focal plane of $\sim 11.8 $ $\rm{ G}$ (see Fig. \ref{fig:HWP}), which gives a required shielding factor $> 59$ for the Cryoperm box surrounding the focal plane. From investigation with commercial partners, this shielding level is easily
attainable with existing material and technologies, but we still plan to verify the requirement of the detector magnetic field tolerance, and a proper electromagnetic simulation to study the magnetic field across the focal plane and the best geometry for the magnetic shield.
\vspace{-5mm}
\subsection{Optics}
\vspace{-3mm}
Ideally, a scaled version of LiteBIRD LFT telescope would allow us to recreate the same conditions we will have during observation. Unfortunately, a cross-dragone mirror configuration with a rotating HWP will require a very complex supporting structure which is difficult to accommodate due to the limited volume of the existing cryostat; therefore, we intend to simplify the optics using a biconvex UHMWPE (Ultra-high-molecular-weight polyethylene) lens with f/\# =3 to match the LFT design and the f/\# of the beam-formers (hemispherical silicon lenslets). We will place the lens after the HWP, and use a parabolic mirror as the first optical element, to collimate the beam through the HWP in order to be able to study its optical properties under simulated working conditions (see Fig. \ref{fig:testbed}). Using a lens requires an anti-reflection coating (\cite{matsumura16}) which we will develop by mechanical machining sub-wavelength structures on the UHMWPE lens surfaces. 
\vspace{-5mm}
\section{Current setup and results}
\vspace{-3mm}
In preparation for the proposed system, we have constructed a simple setup consisting of a few DC SQUIDs (\cite{magnicon}) to study the detector response and noise properties. 

In Fig. \ref{fig:TES1} (left), we show the measured superconducting transition of a TES detector (not LiteBIRD-specific). We performed this measurement in 2 cases: increasing the bath temperature (ramp-up) and a decreasing it (ramp-down). We notice a mismatch in the measured R-T curve between the ramp-up and ramp-down cycles. This effect is likely due to the TES normal resistance (for $T_b>T_c$) acting as an extra heat source, therefore keeping the effective
\begin{figure}[htbp]
    \vspace{-5mm}
    \begin{center}
    \subfloat{{\includegraphics[width=.45\textwidth]{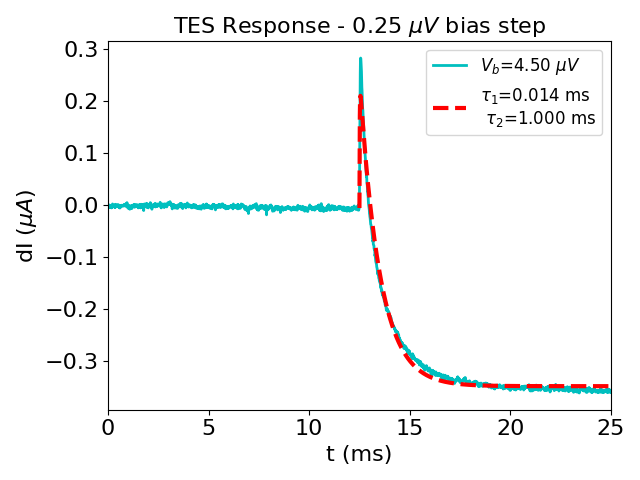} }}%
    \subfloat{{\includegraphics[width=.45\textwidth]{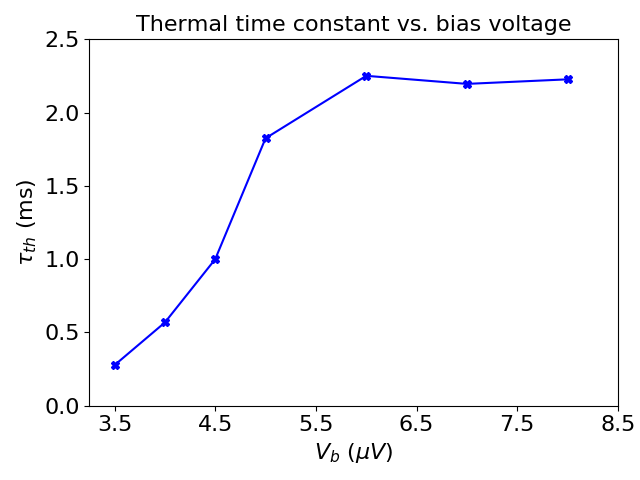} }}%
    \caption{\small{Time constant measurement of a TES detector. {\it Left:} A measurement of the time constant of a TES detector for $\mathcal{L} \sim 1$. The TES steady state output current has been shifted to $0$ $\rm{\mu A}$. We can clearly see a fast rise due to the electrical response of the system and a decay due to the thermal response. {\it Bottom right:} Measured thermal time constant as a function of the bias voltage. (Color figure online.)}} 
    \label{fig:TES2}%
    \end{center}
    \vspace{-5mm}
\end{figure}temperature of the TES higher than the nominal bath temperature during a ramp-down cycle. 

In Fig. \ref{fig:TES1} (center) we show the TES response to a $0.25$ $\rm{\mu V}$ bias-step (\cite{stever19}). For this test we choose a few bias voltage values (in the range 3.5-8 $\mu$V), and we use a signal generator in series with the bias line to input a signal, in this case a slow ($4$ $\rm{Hz}$) square wave.
We can see the stabilization of the response going from high bias voltage (loop gain $\mathcal{L} \sim 0$) to low bias voltage ($\mathcal{L} \gg 1$).
For reference, and to cross-check our correct understanding of the response of the detector, we modelled the response to a bias step; in Fig. \ref{fig:TES1} (right) we show a simulation of the TES response to validate our data. These simulations are performed by numerically solving the linear differential equations governing the thermal and electrical response of a TES to a changing bias voltage (small signal limit), see \cite{irwin05} for details.

In Fig. \ref{fig:TES2}, using the data in Fig. \ref{fig:TES1} (center), we fit the data to extrapolate the thermal time constant of the detector for a each bias voltage value. From the TES response model we expect to find a second time constant ($\sim L/R$) due to the electrical properties of the system. Given an input inductance of 6 nH and the TES resistance $\sim1$ $\Omega$, we expect the value of the electrical time constant to be $\sim 10^{-9}$ s. However, the sampling rate for these measurements is 100 kHz, which limits the possibility to measure such a small value. We find $\tau_1\sim10^{-5}$ s, consistent throughout the measurements and consistent with the sampling rate limit. We find values of $\tau_2$, the thermal time constant of the detector, decreasing for increasing loop gain as expected ($\tau_{th}=C/(G\times (1+\mathcal{L}))$).
As shown in Fig. \ref{fig:TES2}, the thermal time constant of the detector under test decreases from a value of $\sim 2.2$ $\rm{ms}$ to $\sim 0.27$ $\rm{ms}$ for the lowest voltage value, which indicates a loop gain value of $\sim 10$.

We plan to perform more tests with this system, including measurement of noise, saturation power, and sensitivity to external magnetic fields and cosmic rays.
\vspace{-5mm}
\section{Conclusion}
\vspace{-3mm}
We have presented a short summary of the results of the detector tests we are conducting at IPMU in preparation for assembling the proposed testbed polarimeter to study and validate LiteBIRD LFT system and sub-system performances. We believe this is a first but fundamental step to pave the way for full telescope integration, because it will reduce risks, do validate our design approach, and build useful capabilities within the collaboration well ahead of the final delivery.
\vspace{-3mm}
\begin{acknowledgements}
TG acknowledges a Oxford-IPMU joint fellowship for funding his doctoral studies. TG, TM, MH, SLS, YS, NK acknowledge Kavli IPMU supported by World Premier International Research Center Initiative (WPI), MEXT, Japan. This work was supported by JSPS KAKENHI Grant Number JP18KK0083 and by JSPS Core-to-Core Program, A. Advanced Research Networks.
\end{acknowledgements}

\pagebreak

\end{document}